\documentclass[prd,
12pt,
groupedaddress, preprintnumbers,nofootinbib,floatfix]{revtex4-2}
\usepackage[top=3cm, bottom=2.1cm, left=2cm, right=2cm]{geometry}
\pdfoutput=1

\usepackage{comment}

\usepackage{graphicx}
\usepackage{url}
\usepackage[bookmarks, pagebackref=false]{hyperref}
\usepackage[usenames,dvipsnames]{xcolor}
\usepackage{dcolumn}
\usepackage{bm}
\usepackage{bbm}
\usepackage{amsmath,amssymb,amsfonts}
\usepackage{color}
\usepackage{hyperref}
\usepackage[Symbolsmallscale]{upgreek}
\usepackage{dsfont}
\usepackage[all]{xy}
\usepackage{pstricks}
\usepackage{dsfont}%
\usepackage{mathtools}
\usepackage{placeins}
\usepackage{enumerate}
\usepackage{cases}
\usepackage{xspace}
\usepackage{bbold}
\usepackage{cancel} 
\usepackage{slashed}
\usepackage{subfigure}
\usepackage{natbib}
\usepackage{ulem}

\newcommand{\be}{\begin{equation}}
	\newcommand{\ee}{\end{equation}}
\newcommand{\beq}{\begin{eqnarray}}
	\newcommand{\eeq}{\end{eqnarray}}

\newcommand{\bvec}[1]{{\bf{#1}}}
\newcommand{\tsbs}[1]{{\textsubscript{#1}}}

\begin{document}

\title{Energy loss in low energy nuclear recoils in dark matter detector materials}

\author{Sebastian Sassi}
\email{sebastian.k.sassi@helsinki.fi}
\affiliation{Department of Physics, University of Helsinki, 
                      P.O.Box 64, FI-00014 University of Helsinki, Finland}
\affiliation{Helsinki Institute of Physics, 
                      P.O.Box 64, FI-00014 University of Helsinki, Finland}

\author{Matti Heikinheimo}
\email{matti.heikinheimo@helsinki.fi}
\affiliation{Department of Physics, University of Helsinki, 
                      P.O.Box 64, FI-00014 University of Helsinki, Finland}
\affiliation{Helsinki Institute of Physics, 
                      P.O.Box 64, FI-00014 University of Helsinki, Finland}

\author{Kimmo Tuominen}
\email{kimmo.i.tuominen@helsinki.fi}
\affiliation{Department of Physics, University of Helsinki, 
                      P.O.Box 64, FI-00014 University of Helsinki, Finland}
\affiliation{Helsinki Institute of Physics, 
                      P.O.Box 64, FI-00014 University of Helsinki, Finland}                 

\author{Antti Kuronen}
\email{antti.kuronen@helsinki.fi}
\affiliation{Department of Physics, University of Helsinki, 
                      P.O.Box 64, FI-00014 University of Helsinki, Finland}
\affiliation{Helsinki Institute of Physics, 
                      P.O.Box 64, FI-00014 University of Helsinki, Finland}          

\author{Jesper Byggm\"astar}
\email{jesper.byggmastar@helsinki.fi}
\affiliation{Department of Physics, University of Helsinki, 
                      P.O.Box 64, FI-00014 University of Helsinki, Finland}
\affiliation{Helsinki Institute of Physics, 
                      P.O.Box 64, FI-00014 University of Helsinki, Finland}     

\author{Kai Nordlund}
\email{kai.nordlund@helsinki.fi}
\affiliation{Department of Physics, University of Helsinki, 
                      P.O.Box 64, FI-00014 University of Helsinki, Finland}
\affiliation{Helsinki Institute of Physics, 
                      P.O.Box 64, FI-00014 University of Helsinki, Finland}         
\author{Nader Mirabolfathi}                      
\email{mirabolfathi@physics.tamu.edu} 
\affiliation{Department of Physics and Astronomy and the Mitchell Institute for Fundamental Physics and Astronomy,
Texas A\&M University, College Station, TX 77843, USA}

\begin{abstract}
\noindent
Recent progress in phonon-mediated detectors with eV-scale nuclear recoil energy sensitivity requires an understanding of the effect of the crystalline defects on the  energy spectrum expected from dark matter or neutrino coherent scattering. 
We have performed molecular dynamics simulations to determine the amount of energy stored in the lattice defects as a function of the recoil direction and energy. This energy can not be observed in the phonon measurement, thus affecting the observed energy spectrum compared to the underlying true recoil energy spectrum. We describe this effect for multiple commonly used detector materials and demonstrate how the predicted energy spectrum from dark matter scattering is modified.
\end{abstract}
\preprint{HIP-2022-13/TH}
\maketitle

\section{Introduction}

The cosmological and astronomical evidences for the existence of dark matter (DM) that comprises vast majority of the mass content of the universe are abundant~\cite{Planck:2018vyg, Bertone:2004pz}. Among dark matter candidates, many models predict particles 
that can elastically  scatter off the electrons or nuclei releasing a measurable energy in the detectors. Due to the very low energy recoils expected from galactic DM halo, the interaction of DM particles with detector nucleus is particularly interesting. This is because at low momentum transfers, the particles are expected to coherently interact with the nucleons in the nucleus increasing the cross-section of interaction by $A^2$ where $A$ is the atomic number of the target material. 
The attempts of such direct detection of DM scattering on nuclei have in the past mostly focused on the DM mass range $m_{\rm DM}\gtrsim 1$ GeV (we use natural units $\hbar=c=1$), and the most stringent exclusion limits for DM nucleon scattering have been obtained for $m_{\rm DM}\gtrsim 10$ GeV~\cite{XENON:2018voc}. 

Very low threshold detectors are needed to measure the small energies from DM interactions with nuclei for DM masses in the sub-GeV range. Recently several groups ~\cite{CDMS:CPD,CRESST:2019jnq,DAMIC:2020cut,EDELWEISS:2019vjv,Strauss:2017cam}, have developed detectors with detection thresholds in the $\mathcal{O}(10)$ eV recoil energy range. Among those technologies, the phonon-mediated detectors are particularly interesting because their energy measurement indicates the true recoil energy. This is in contrast to the ionization or scintillation detectors wherein the true recoil energy should be corrected for the ionization or scintillation yield~\cite{CDMSNR}.  
Currently these experiments observe an event rate in excess of the anticipated background~\cite{Proceedings:2022hmu}. 
Until the origin of this excess is clarified, it provides an obstruction for reaching the necessary sensitivity of the experiment to a DM signal.

At recoil energies below a few hundred eV it becomes necessary to understand the detailed response of the target material to the DM scattering. Recent developments on the theoretical description of matter-effects in direct detection of light dark matter include \cite{Kahn:2021ttr,Mitridate:2022tnv,Campbell-Deem:2022fqm} and the references therein.
For nuclear recoils, the observed recoil spectrum in phonon based detectors can be influenced by the creation of crystal defects in the target material: part of the true recoil energy is stored in the defect, and will not reach the phonon detector, resulting in a lower observed energy for the event. To understand and interpret the observed spectrum in terms of the physical processes responsible for the recoil events, this effect needs to be accounted for~\cite{Kadribasic:2020pwx}. 
For this purpose, the energy stored in the defects must be known as a function of the recoil direction and recoil energy.

In this paper we report the results from molecular dynamics simulations performed for multiple commonly used or proposed detector materials. These are sapphire (Al\tsbs{2}O\tsbs{3}), silicon carbide (SiC), tungsten carbide (WC), diamond (C), silicon (Si), germanium (Ge) and tungsten (W). Using these results, we then describe the resulting modifications in the expected dark matter recoil spectrum in phonon based detectors. We have proposed a method utilizing this effect 
to identify the origin of the low energy excess events observed in multiple experiments in \cite{Heikinheimo:2021syx}.

The paper is organized as follows: In section~\ref{sec:MDsection} we present the setup and results of the molecular dynamics simulations. Then, in section~\ref{sec:DMrecoils} we explain how the crystal defects affect the observed spectrum of DM recoils emphasizing the dependence on the target material. In section~\ref{sec:checkout} we present our conclusions and outlook for further work.

\section{Molecular dynamics simulations}
\label{sec:MDsection}

\subsection{Simulation setup}

The molecular dynamics simulations for Al\tsbs{2}O\tsbs{3}, SiC, WC, and W were done using the LAMMPS molecular dynamics simulator \cite{LAMMPS}, while the simulations for C, Si and Ge were done previously using the PARCAS code\cite{Nor97f,PARCAS}. Otherwise the principles for these simulations are the same as described below, following general principles previously used to examine threshold displacement energies in materials \cite{Cat94,Nor97f,Kadribasic:2017obi}. The simulation parameters and potentials used for each material are listed in table \ref{tab:params}.

The simulation region was constructed from orthogonal unit cells such that the region would have roughly the same size in each dimension and would contain $O(10^3)$ atoms. For SiC and WC, that have a hexagonal conventional unit cell, an orthogonal unit cell twice the size of the hexagonal cell was used. The boundary conditions were periodic in each direction. The simulations were done at a temperature of 40 mK. To achieve this, for each material the simulation box was first allowed to reach an equilibrium state by running it for 1 ps under temperature and pressure control at the target temperature of 40 mK under no external pressure. The final configuration of this simulation would then be used as the starting configuration of all the recoil simulations for that material. We note that in classical molecular dynamics, the quantum mechanical zero-point vibrations of atoms are not included in the description of atom motion. However, we have previously found by comparing 40 mK results with higher temperature ones that this does not affect the damage production probability because the recoil kinetic energies $> 10$ eV are several orders of magnitude higher than the meV thermal vibration energies \cite{Kadribasic:2017obi}.

To let kinetic energy dissipate in the recoil simulations, the simulation box was divided into two regions: an interior where the recoils would take place, and a border region under temperature control at 40 mK to dissipate thermal energy into the surroundings. The thickness of the border region was set to 6.0 angstroms in all simulations.

To simulate nuclear recoils over a given range of recoil energies, for each atom inside the unit cell closest to the center of the simulation box random recoil directions were sampled uniformly on the sphere. For a given direction, for each recoil energy in the simulated range the energy and direction would be used to set the velocity of the atom to simulate increasingly energetic recoils in the direction. The numbers of simulated directions vary for the different materials, but all the data presented here is based on $\gtrsim 1000$ recoil directions. After the velocity of the recoiling atom was set, the simulation was run for a fixed amount of time, depending on the target material, to let the system return as close to an equilibrium as viable. This time was primarily constrained by practical considerations of execution time. A sufficient simulation time was determined by running longer test simulations ($>5$ ps) at different recoil energies and observing how long it took for the system's potential energy to fall to within $<1$ eV of its final value. The size of the time step for each material was chosen based on similar considerations of stability of the system's energy under changes in size of the time step.

As alluded to above, the energy lost to defects was based on calculating the difference in the potential energy of the system at the start of the simulation before the recoil, and at the end of the simulation. Since the permanent energy loss in these materials should be associated with creation of at least one Frenkel pair, a consistency check was performed on the occupation numbers of each Wigner--Seitz cell (for non-Bravais lattices this is the Voronoy polyhedron surrounding each atom position \cite{Ashcroft-Mermin}) in the simulation box, with occupation number differing from one indicating a defect. Although there are numerous ways to analyze defects in crystals, this Wigner-Seitz cell/Voronoy polyhedron approach has the advantage of not  involving any free parameters, and not counting bond order defects that likely are metastable \cite{Nor97f}.
If no defects were found, a value of zero would be recorded for the energy loss. Else the potential energy difference at the start and end of the simulation was recorded.

\begin{table}
    \begin{tabular}{llll}
        \hline
                            & Al\tsbs{2}O\tsbs{3}                    & SiC                   & WC                        \\\hline
        Unit cell config.   & $8\times 5\times 3$       & $5\times 9\times 3$   & $10\times 6\times 10$     \\
        Atoms per unit cell & 60                        & 16                    & 4                         \\
        Time step (ps)      & 0.0005                    & 0.0005                & 0.00025                   \\
        Simulation time (ps)& 4.0                       & 4.0                   & 3.2                       \\
        Potential           & Vashishta et al. \cite{Vashishta2008}& Gao--Weber \cite{Gao2002}& Juslin et al. \cite{Juslin2005}   \\\hline
                            & C                & Si                   & Ge                        \\\hline
        Unit cell config.   & $8\times 8\times 8$       & $8\times 8\times 8$   & $8\times 8\times 8$     \\
        Atoms per unit cell & 8                        & 8                    & 8                         \\
        Time step (ps)      & Adaptive                    & Adaptive                & Adaptive                  \\
        Simulation time (ps)& 20.0                       & 20.0                   & 20.0                       \\
        Potential           & Erhart \cite{Erh04},  &  Stillinger--Weber \cite{Sti85}& Modified Stillinger--Weber \cite{Nor97f}   \\
        & Tersoff--Nordlund \cite{Ter89,Nor96} & & \\
        \hline
                            & W                 &                   &                        \\\hline
        Unit cell config.   & $10\times 10\times 10$       &    &     \\
        Atoms per unit cell & 2                        &                    &    \\
        Time step (ps)      & 0.00009                    &         &                   \\
        Simulation time (ps)& 4.2                       &                    &                        \\
        Potential           &  Derlet--Björkas \cite{derlet_multiscale_2007,Bjorkas2008} &  &  \\\hline

    \end{tabular}
    \caption{Parameters and potentials used in the simulations. The simulations for C, Si and Ge were done previously using the PARCAS code\cite{Nor97f,PARCAS} that employs an adaptive time step algorithm in radiation effect calculations \cite{Nor94b}. }
    \label{tab:params}
\end{table}

\subsection{Results}

\begin{figure}
    \centering
    \includegraphics{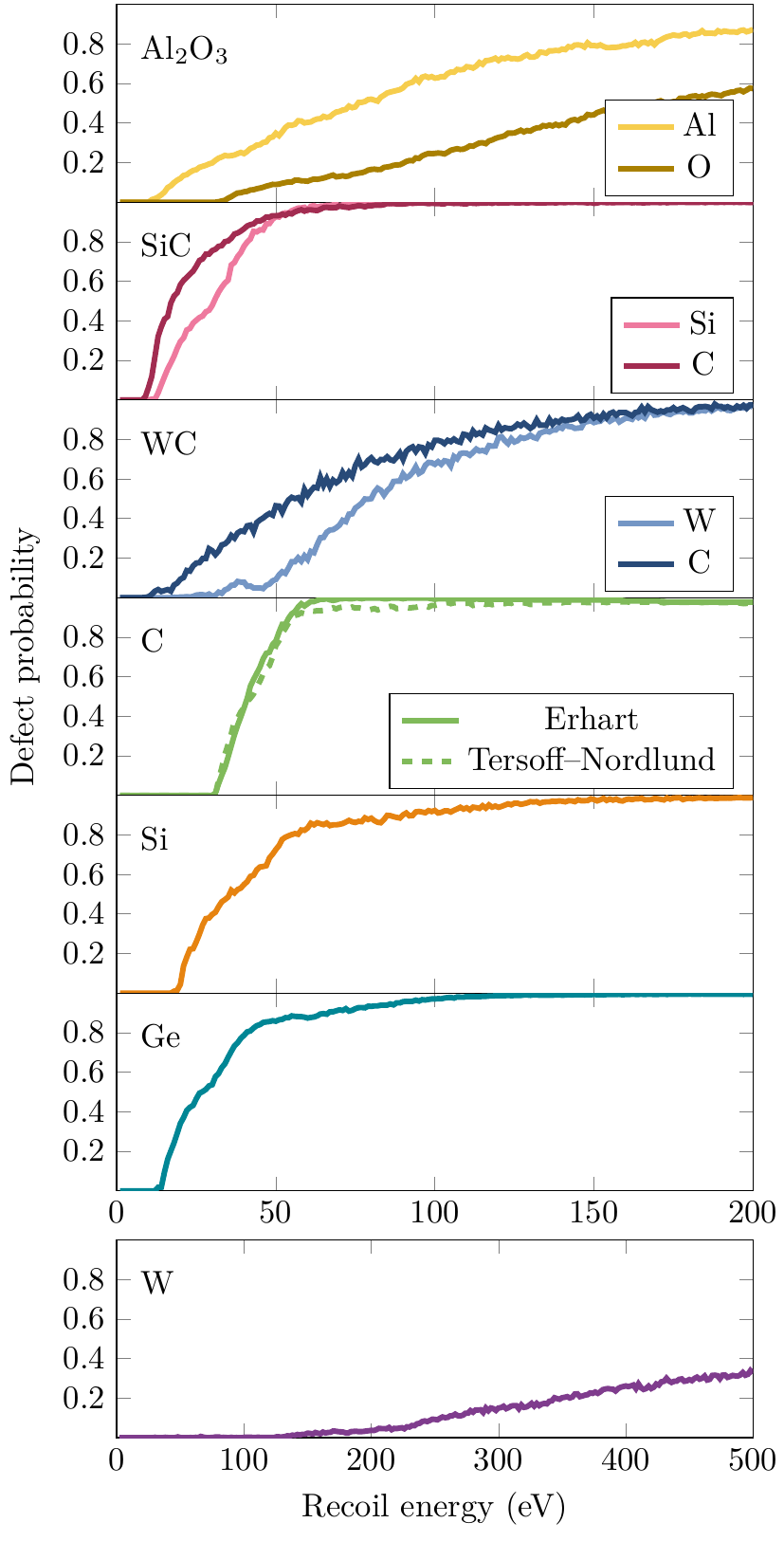}
    \includegraphics{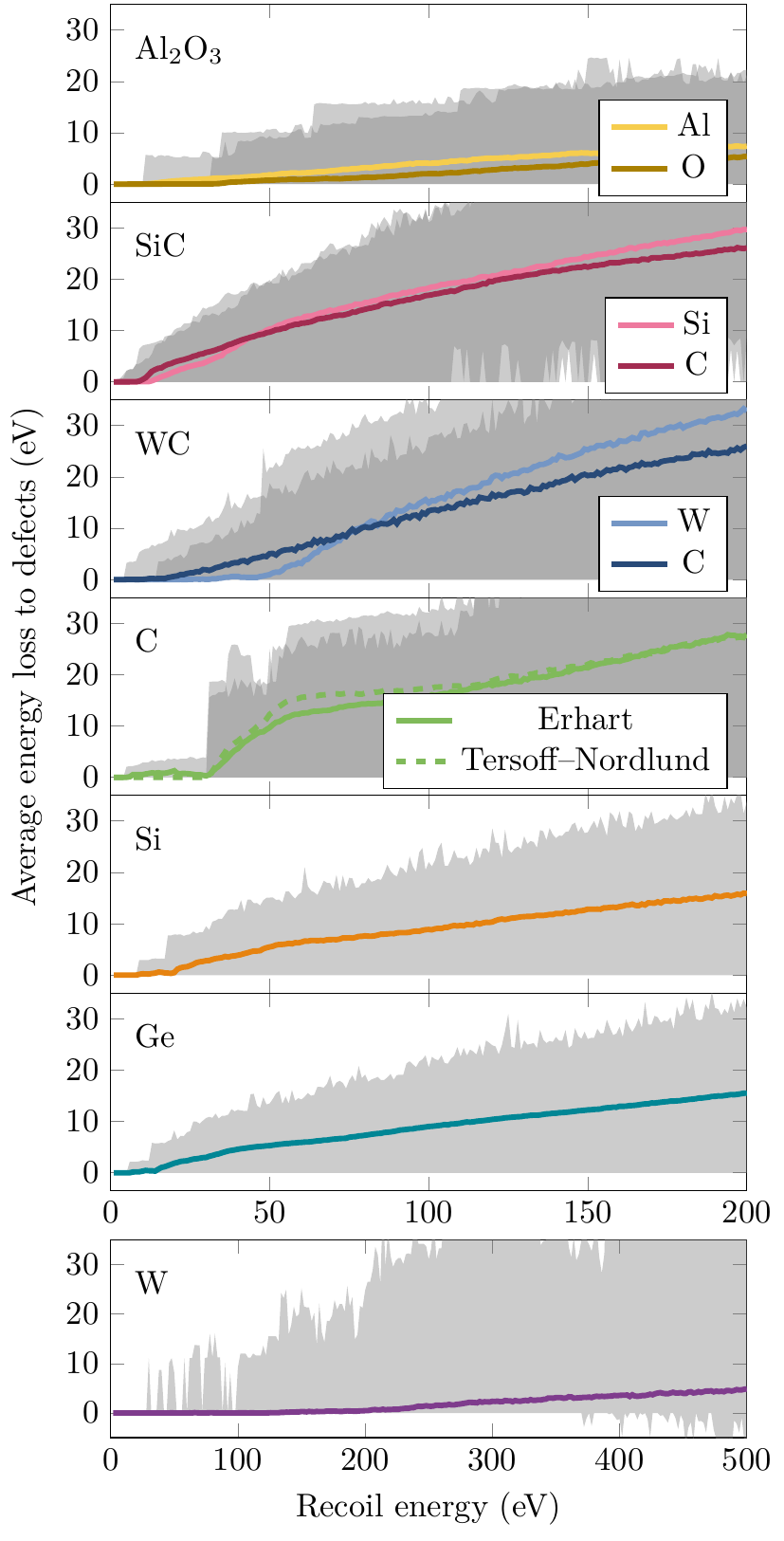}
    \caption{Left: probability of defect creation from a nuclear recoil as a function of recoil energy in the simulated materials. Right: the solid lines show the average energy loss in the materials, and the grey bands show the range of energy loss. The minimum defect creation energies for each material are listed in table \ref{tab:thresholds}. Figures \ref{fig:elossprob1} and \ref{fig:elossprob2} show the spread of energy loss in more detail.}
    \label{fig:defprob_avgeloss}
\end{figure}

\begin{table}
    \begin{tabular}{lll}
        \hline
        Material & Recoil atom & $E_\text{t,min}$ (eV)\\\hline
        Al\tsbs{2}O\tsbs{3} & Al & 11\\
        Al\tsbs{2}O\tsbs{3} & O & 31\\
        SiC & Si & 11\\
        SiC & C & 9\\
        WC & W & 18\\
        WC & C & 9\\
        C--Erhart & & 31\\
        C--TN & & 31\\
        Si & & 18\\
        Ge & & 13\\
        W & & 30\\\hline
    \end{tabular}
    \caption{Minimum defect creation threshold energies $E_\text{t,min}$ for the simulated materials.}
    \label{tab:thresholds}
\end{table}

We find that there exists significant variance in the energy loss characteristics between the simulated materials. On the left hand side of figure \ref{fig:defprob_avgeloss} the defect probabilities, defined here as average over the sampled directions, at different recoil energies are shown. It is noteworthy that for pure elemental C, simulations were performed with two different interatomic potentials. Throughout the results, as seen in figure \ref{fig:defprob_avgeloss}, it will be evident that there is little difference in defect creation and energy loss between the different potentials, giving some confidence that these results are not highly sensitive to the details of the model for interatomic forces.

In the single-element materials and in SiC we see that the defect creation probability from nuclear recoils generally rises steeply after the initial minimum energy threshold for defect creation is reached. Although in SiC the defect probability for C recoils rises later, the probability for both elements quickly approaches unity above recoil energies of 50 eV. In sharp contrast, in Al\tsbs{2}O\tsbs{3}, the defect creation probability only rises gradually, with significant likelihood of no defect creation even at recoil energies of 200 eV for both Al and O recoils, and O recoils remaining significantly less likely than Al recoils throughout the energy range. WC is an interesting case because of the large difference in masses of the W and C atoms. Here there is a significant difference in the initial threshold for defect creation between the atoms, which is likely due to the aforementioned mass difference.
Regardless of the initial difference, around 200 eV, the defect probability from both types of recoils again approaches unity.

 The defect formation probability in pure W differs strongly from the other materials considered in this work. This is to be expected, since densely packed elemental metals have a very strong athermal damage recombination effect \cite{Dia87,Nor18}. This causes the damage production in common metals to be much less efficient than in covalently bonded materials such as Si and Ge \cite{Nor97f}. Due to this effect, both the defect formation probability and permanent energy loss is much lower than in the non-metallic materials considered in this study (Fig. \ref{fig:defprob_avgeloss}).

\begin{figure}
    \centering
    \includegraphics{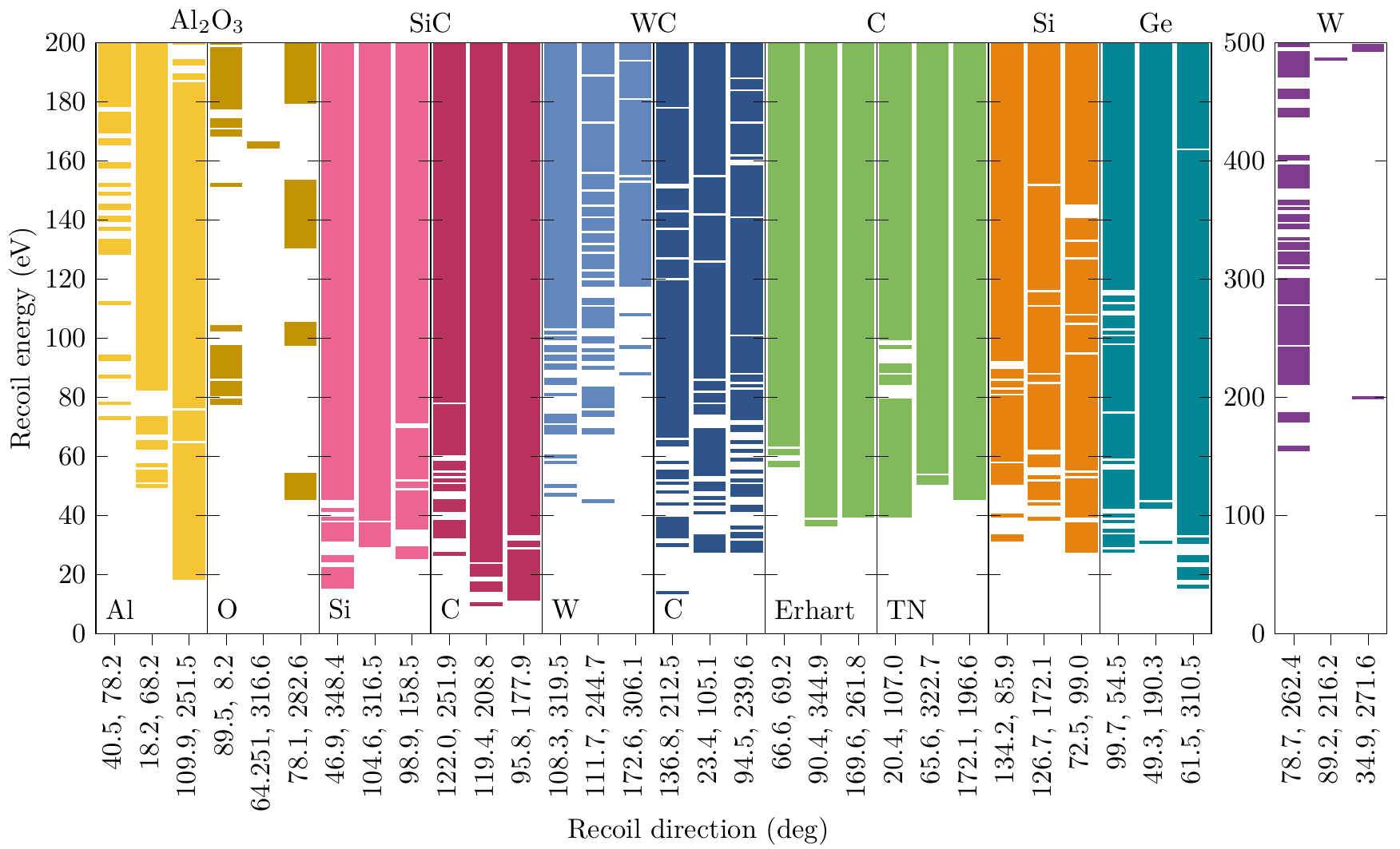}
    \caption{Illustrative sample of directions demonstrating typical defect creation in the simulated materials. Energies for which a defect was created have been colored in, while energies that produced no defects have been left blank. The direction of each recoil is shown below in degrees with the first angle being the colatitude and the second the azimuthal angle. Note that the choice of directions has no special significance.}
    \label{fig:defbin}
\end{figure}

The behavior of the defect formation probability as a function of recoil energy could in principle result from directional variation in the initial defect creation threshold, or recombination of defects moments after their creation. Indeed, we find both effects at play. While there is directional variation in the materials, there is significant variation between materials in how likely defects are to form after the threshold recoil energy for defect creation has been exceeded. Figure \ref{fig:defbin} shows defect creation and recombination trends for an illustrative sample of directions for each material, where the recoil energies at which defects were created have been colored. The samples have been picked from the full simulated sample of random directions to showcase typical behavior in the given material. Here it is seen that for elemental C, as well as SiC, at recoil energies above 100 eV defects form almost always. Si and Ge display a little more recombination, which is reflected in the slower rise of the defect creation probability. For Al\tsbs{2}O\tsbs{3} in turn, especially in O recoils, defects are highly likely to recombine even at the maximum simulated recoil energies of 200 eV. 

\begin{figure}[t]
\centering
    \includegraphics{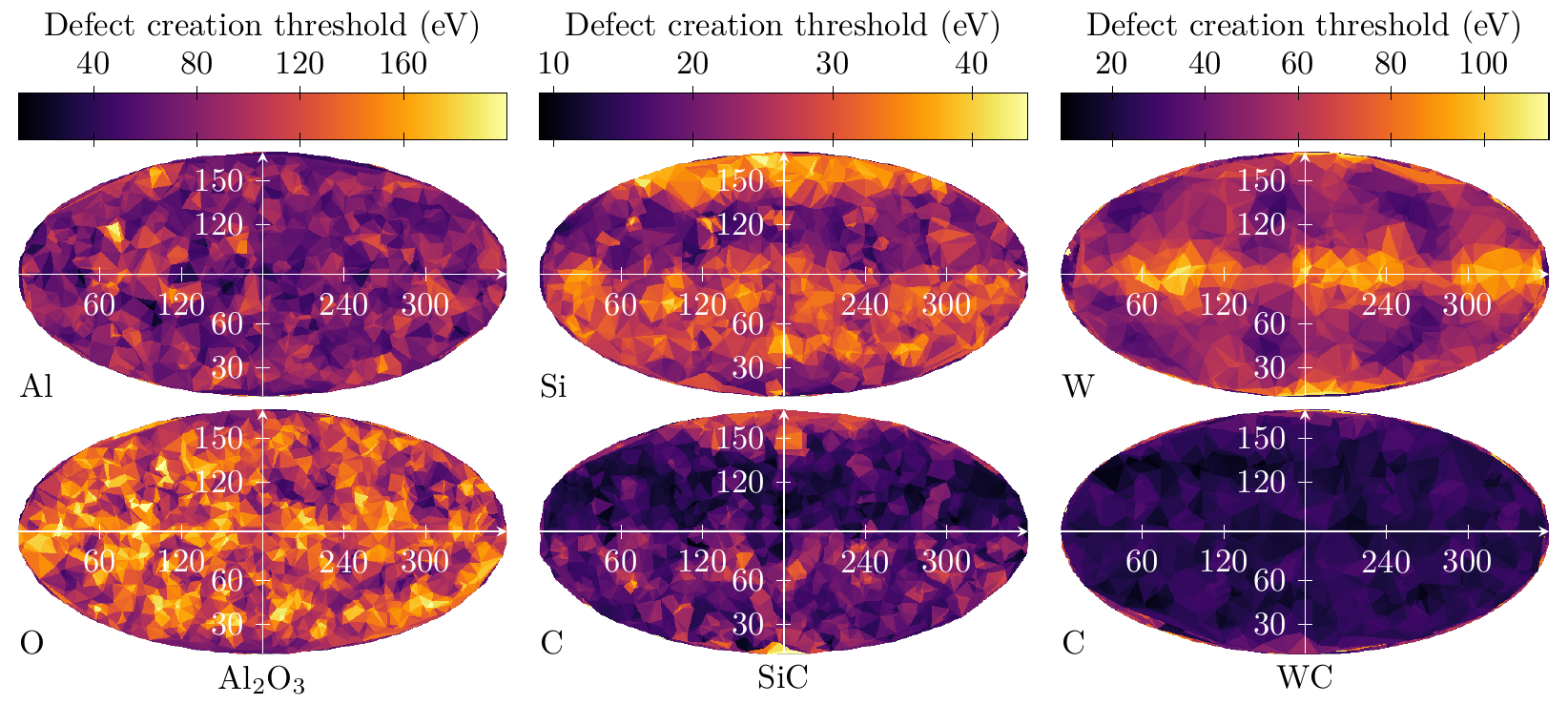}
	\includegraphics{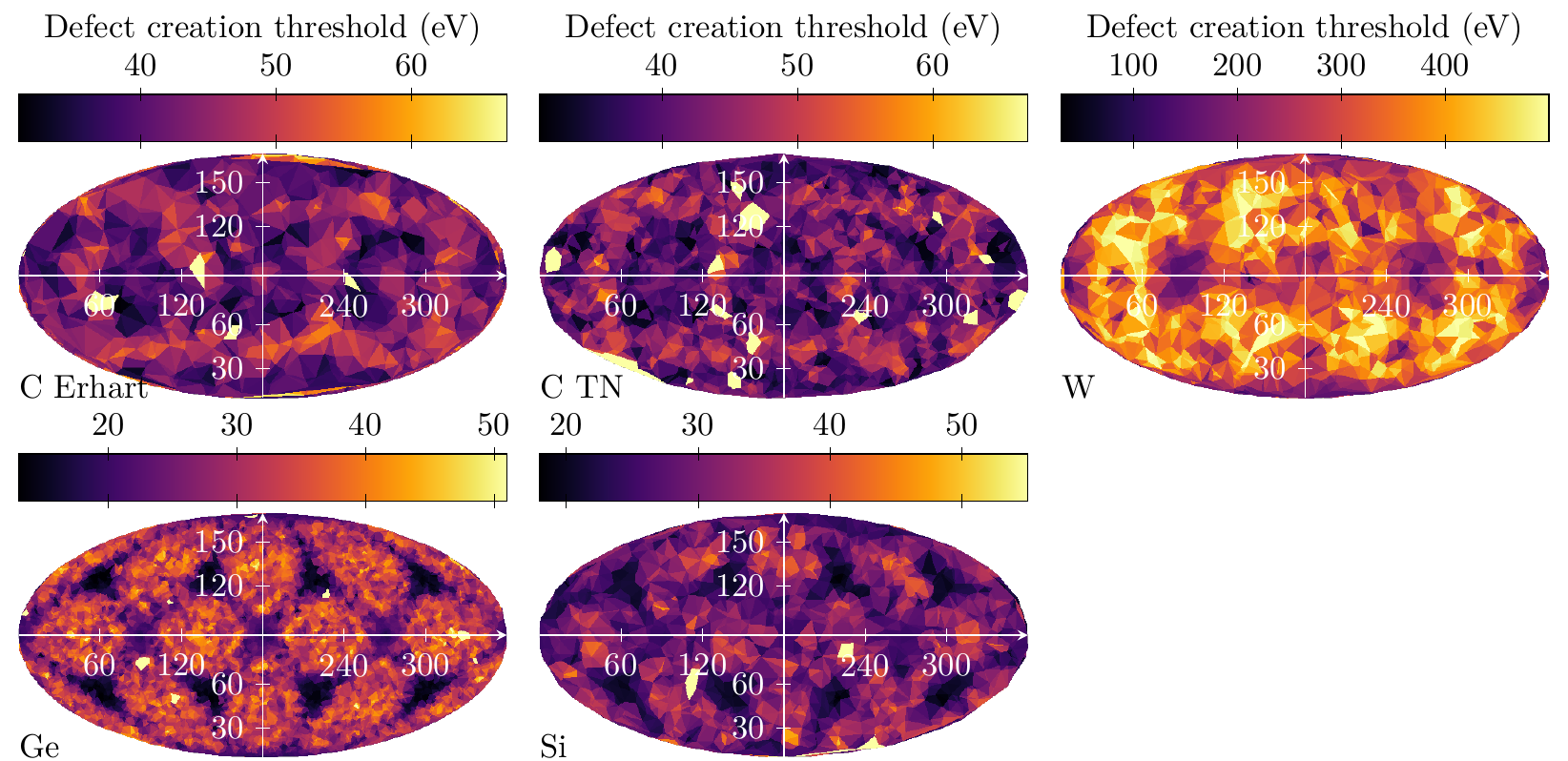}
	\caption{Mollweide projection of the angular distribution of defect creation thresholds in the materials. The top six plots show the compound materials for each recoil atom. The bottom five plots show the single-element materials. Directions where no defect was produced at any energy have been mapped to the highest simulated recoil energy. Because of this, in order to show the directional anisotropy, the maximum value of the color map is not the highest threshold energy, but has been chosen to be the 99th percentile threshold energy. It is also worth noting that the threshold energy here is merely the lowest energy at which we observed a defect in our simulations. Due to the probabilistic nature of defect production around the threshold, these values may not correspond to the true threshold in the sense of, e.g., the recoil energy above which 50\% of recoils lead to defects. Therefore while in the materials where recombination of defects is rare (such as Ge) the distribution is likely fairly accurate, in materials with lots of recombination such as Al\tsbs{2}O\tsbs{3} there is likely a significant amount of noise.}
	\label{fig:directional}
\end{figure}

Although directional trends are not a focus of this study, we have produced plots of the angular distribution of defect creation thresholds in figure \ref{fig:directional} for benefit of the reader with the caveat that due to the recombination trends shown in figure \ref{fig:defbin} some of these may be dominated by noise. A detailed description of the directional behavior of defect creation and energy loss requires further research.

\begin{figure}
    \centering
    \includegraphics{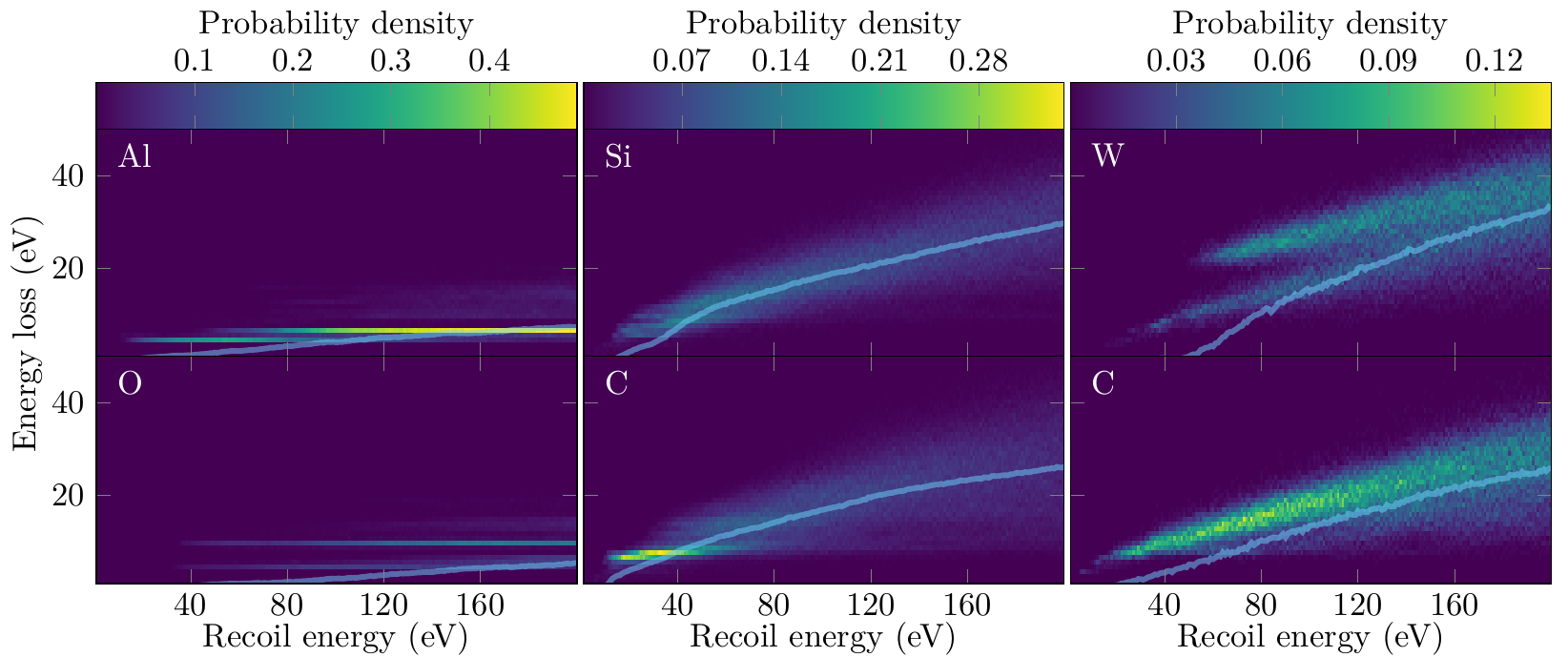}
    \caption{Probability distribution of energy loss parametrized by recoil energy for Al\tsbs{2}O\tsbs{3}, SiC, and WC. The energy loss has been binned with 1 eV intervals. The solid line shows the average energy loss. Note that the 0 eV bin (no defects) is included in the probability density, but has been left off the figure to focus on the distribution of energy loss when defects are produced. As a consequence the average energy loss trails below the bands in the figure when the probability of defect production is low.}
    \label{fig:elossprob1}
\end{figure}

It is evident that in a material like Al\tsbs{2}O\tsbs{3} with significant recombination, the initial defect creation threshold for a given direction can be difficult to determine. Indeed, some of the simulated directions not shown here show no defects in the whole simulated recoil energy range, but it is clearly wrong to conclude that 200 eV of recoil energy is not sufficient to create defects. Even in the less extreme cases significant recombination obscures the directional variation in the defect creation threshold energy, making it potentially seem more significant than it actually is. Therefore for the materials with significant degree of recombination, this data may not be useful for showcasing directional variation in the defect creation threshold.

On the right hand side of figure \ref{fig:defprob_avgeloss} the direction averaged energy loss for each material is shown as a function of recoil energy. The trends here largely reflect the trends seen in defect creation probability on the left hand side of figure \ref{fig:defprob_avgeloss}, although here the fact that the amount of defects and therefore the energy lost to defects also rises as a function of recoil energy linearizes the trends. Hence Al\tsbs{2}O\tsbs{3} generally has very little energy loss, and because the average defect creation threshold was less pronounced, so is the threshold for energy loss. The average energy loss shows the effect of the different amounts of energy stored in defects in different materials. Although SiC and elemental C, Si, and Ge behave roughly similarly in the defect creation probability, Si and Ge have significantly lower average energy loss due to defects in the latter materials generally storing less energy. Here a minor difference is also seen between the two different potentials for elemental C. In WC, we see the disparity between the W and C recoils pronounced, and while the threshold in energy loss for C is damped, showing a very linear rise, W retains a more pronounced threshold. It is interesting to note in the context of WC that if the scattering is coherent, and so the cross section of the recoil process is proportional to the squared mass number $A^2$, then the the rate of C recoils will be very small compared to the rate of W recoils. Consequently, a recoil experiment measuring energy loss should mainly see the energy loss curve of W recoils.
\begin{figure}
    \centering
    \includegraphics{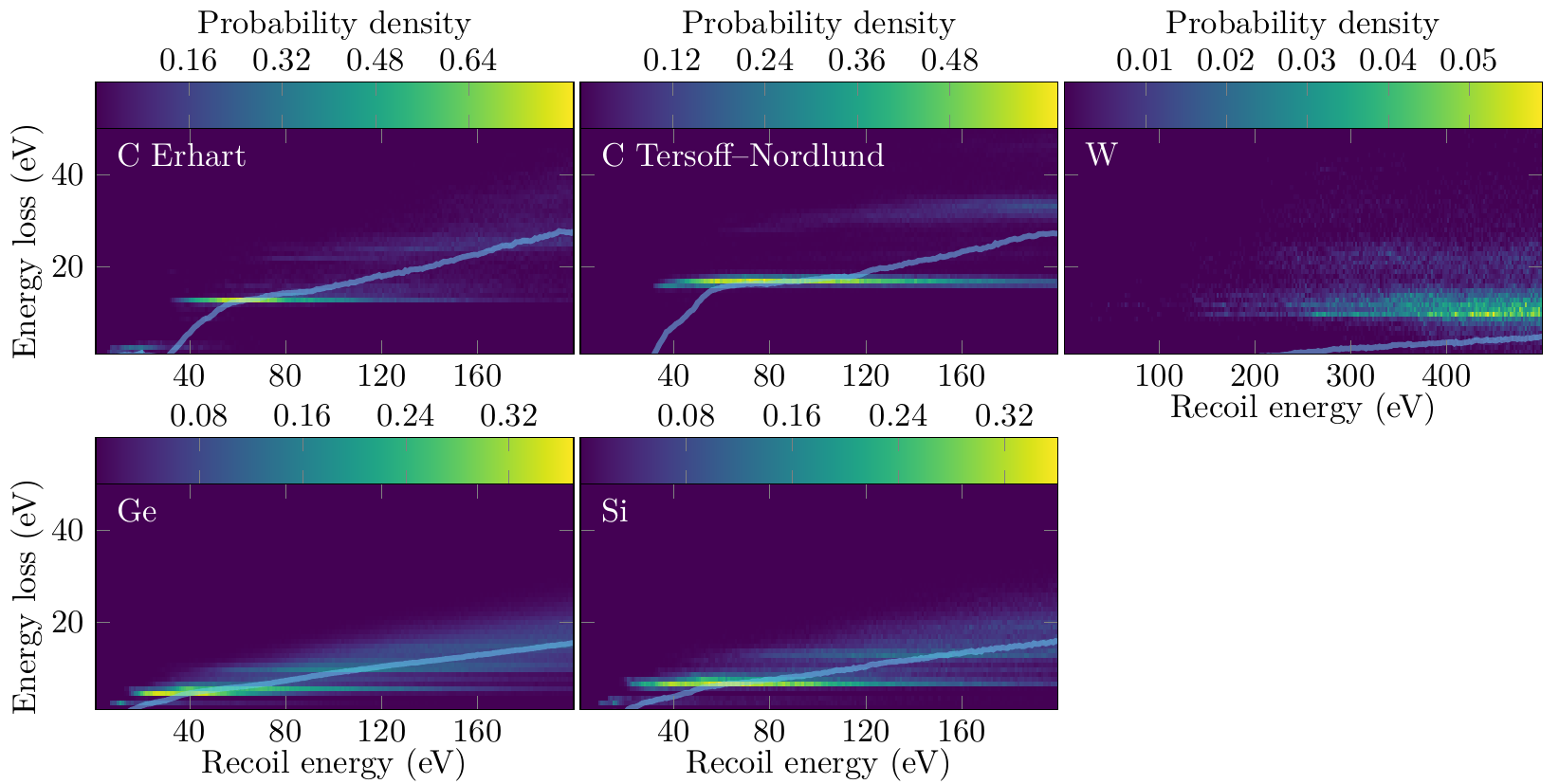}
    \caption{Probability distribution of energy loss parametrized by recoil energy for C with Erhart and Tersoff--Nordlund potential, Ge, Si, and W. See caption of figure \ref{fig:elossprob1} for details.}
    \label{fig:elossprob2}
\end{figure}

Figures \ref{fig:elossprob1} and \ref{fig:elossprob2} show a more detailed picture of the energy loss. Here the probability density of energy loss is plotted at various recoil energies (i.e. adding the probabilities over the energy loss for a given recoil energy gives unity). These figures show the above mentioned differences in energy stored in defects. For elemental C, at low recoil energies, most of the generated defect configurations store around 10--15 eV of energy with nothing below that, whereas for Si, Ge, and SiC the lowest amount of energy stored is around 5 eV. As the recoil energy increases, the distribution of energy loss tends to spread out.
This spreading out is particularly pronounced in the covalently bonded materials SiC and WC, that have fairly complex and open crystal structures. In such materials, there are many possibilities for different metastable disordered defect configurations, which lead to a large variation in the energy loss to defects. Conversely, the very hard C in the diamond structure has also high resistance to radiation damage and hence less variation. The small variation in Al\tsbs{2}O\tsbs{3} is likely due to the strong ionic character in its bonding reducing the number of possible defect configurations (the Coulombic forces between like charges essentially hinder atoms of the same type to remain close to each other). In W, there is also some spread, but note that this spread is at much higher energies than in the other materials due to the recombination effect mentioned above. This spread can be understood in terms of occasional formation of extended defects with complex geometry \cite{Mar12c,Liu13a}.

\section{Recoil spectrum for DM}
\label{sec:DMrecoils}

Because the energy loss depends strongly on the recoil direction, we first compute the expected DM-nucleus recoil event rate as a function of direction and energy, given by 
\be
\frac{dR}{dEd\Omega_q}=\frac{\rho_0 A^2 F(E)\sigma_{\rm DM n}}{4\pi m_{\rm DM} \mu_{\rm DM n}^2}\hat{f}(v_{\rm min},\hat{\bvec{q}}),
\label{eq:dRdEdOmega}
\ee
where $\mu_{\rm DM n} = m_{\rm DM} m_{\rm n}/(m_{\rm DM}+m_{\rm n})$ is the reduced mass of the DM--nucleon system, $A$ is the mass number of the target nucleus, $F(E)$ is the nuclear form factor and $\rho_0$ is the local DM density. For simplicity we have assumed a spin-independent interaction, characterized by the DM-nucleon scattering cross section $\sigma_{\rm DM n}$. The dependence on the recoil direction, parametrized by the recoil momentum unit vector $\hat{\bvec{q}}$, is in the Radon transform of the DM velocity distribution $\hat f$ defined as
\be
\hat{f}(v_{\rm min},\hat{\bvec{q}}) = \int\!\! f(v)\delta(\bvec{v}\cdot\hat{\bvec{q}}-v_{\rm min})\,d^3v,
\ee
where $v_{\rm min} = \sqrt{m_N E/(2\mu_{{\rm DM} N}^2)}$ is the minimum velocity of the DM particle in the lab frame required to excite a nuclear recoil with energy $E$ for a nucleus with a nuclear mass $m_N$ and $\mu_{{\rm DM} N}$ is the DM-nucleus reduced mass.
For simplicity we assume here the standard halo model for the velocity distribution of DM
\be
f_{\rm SHM}(v) = \frac{1}{N_{\rm esc}}\frac{1}{(2\pi\sigma_v^2)^\frac{3}{2}} e^{-\frac{v^2}{2\sigma_v^2}}\Theta(v_{\rm esc}-v),
\ee
where $\sigma_v$ is the velocity dispersion and $v_{\rm esc}$ is the escape velocity. The analytical formulas of the Radon transform $\hat{f}_{\rm SHM}$ shifted to the laboratory frame and integrated over recoil energy can be found in~\cite{Heikinheimo:2019lwg}.

To obtain the observed recoil spectrum in a phonon measurement after the energy loss, we perform a simulation as follows: first, the recoil direction surface (unit sphere) is divided into solid angle bins corresponding to the directions for which the MD simulations were performed as described in section \ref{sec:MDsection}. Each solid angle bin is then further divided into energy bins using 1~eV intervals, for which we find the DM event rate using (\ref{eq:dRdEdOmega}). This procedure results in a binned underlying event rate, reflecting the energy deposited by the DM scattering events. We then sample this underlying event rate, and for each sampled event with  recoil energy $E_{\rm r}$ and recoil direction $\hat{\bvec{q}}$ we find the observed energy as
\be
E_{\rm obs} = E_{\rm r} - E_{\rm loss}(E_{\rm r},\hat{\bvec{q}}) + E_\sigma,
\label{eq:eloss}
\ee
where $E_{\rm loss}(E_{\rm r},\hat{\bvec{q}})$ is obtained from the MD simulations and $E_\sigma$ is a random number drawn from a Gaussian distribution with zero mean and standard deviation $\sigma$ representing the energy resolution of the detector, for which we use $\sigma = 3$~eV. The result after this sampling is the binned observed recoil rate, which we then sum over the solid angle bins as the detector is not capable of observing the recoil direction, to obtain the observed recoil spectrum $dR/dE_{\rm obs}$.

\begin{figure}
	\centering
		\includegraphics{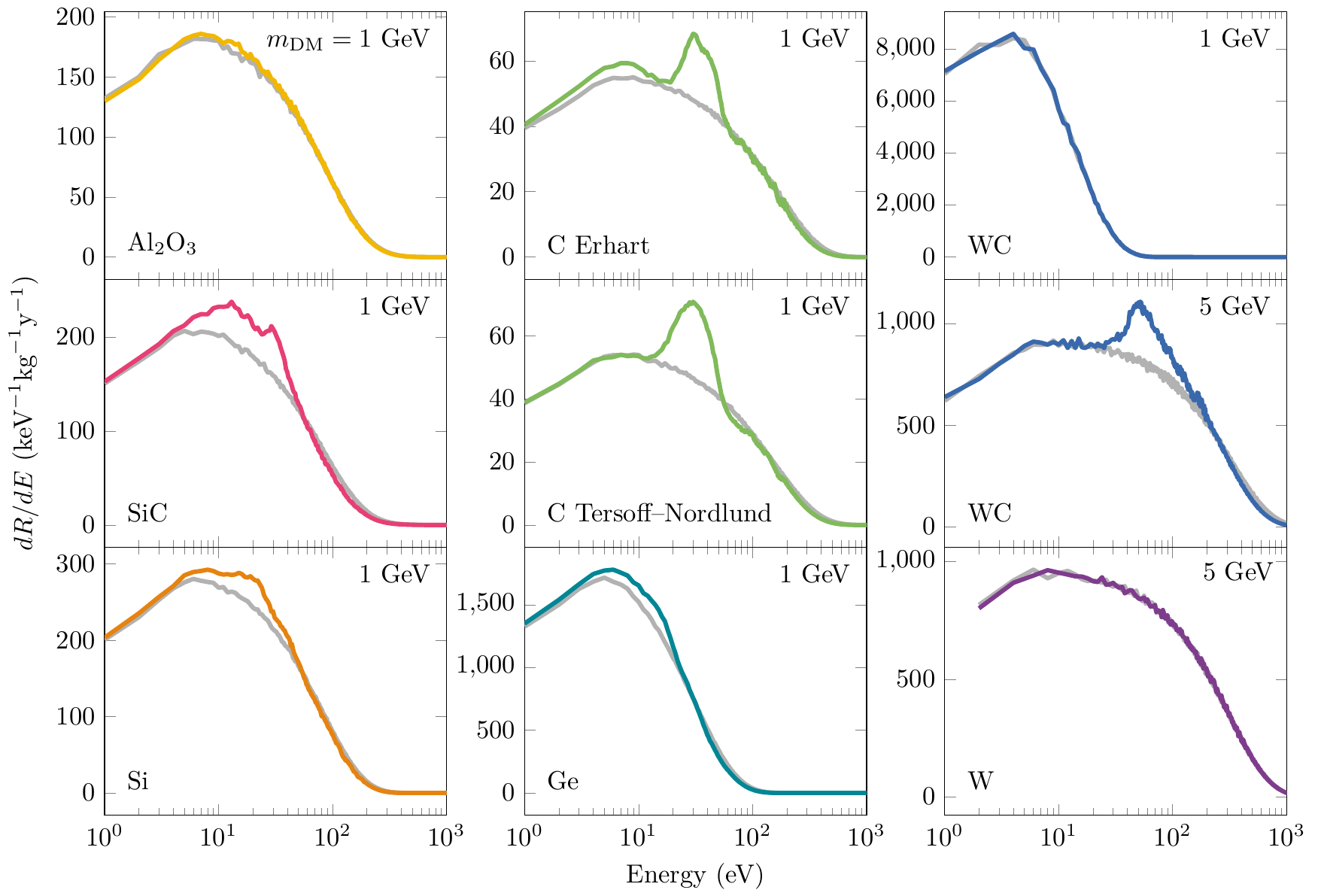}
		\caption{Impact of energy loss on the DM differential event rate $dR/dE_{\rm rec}$. For each material the grey line is the rate in that material without energy loss, and the colored line is the rate after energy loss. The dark matter mass is shown in the top right corner.}
		\label{dRdE}
	\end{figure}

The resulting recoil spectrum is shown in figure \ref{dRdE}, for a 1~GeV DM particle (and 5~GeV for WC and W) in the simulated detector materials, normalized to DM-nucleon cross section $\sigma_{\rm DMn} = 10^{-42}\, {\rm cm}^2$.
The colored curves show the result with the computed energy  loss $E_{\rm loss}$, while the grey curves correspond to the case without the energy loss effect
i.e. setting $E_{\rm loss}=0$ in eq.~\eqref{eq:eloss}.

As seen from the figure, the sharp defect creation threshold in diamond results in a prominent peak in the observed recoil energy around $E_{\rm obs}\sim 30$ eV. For tungsten carbide the energy loss curve is sharp for tungsten nuclei but smooth for carbon nuclei. However, due to the $A^2$ factor in (\ref{eq:dRdEdOmega}) the DM recoil rate in tungsten carbide is completely dominated by tungsten recoils, and therefore the sharp threshold results in a prominent peak feature. Because of the heavy nuclear mass of tungsten, the peak is only visible for DM mass above 1 GeV. Figure \ref{dRdE} shows the differential event rate in WC for 1 GeV and 5 GeV DM particles, and a clear peak is visible in the latter. As discussed in \cite{Heikinheimo:2021syx}, the presence or absence of the peak in the observed recoil spectrum could be used to determine if the observed events are indeed due to nuclear recoils. For this purpose it is important that the peak is not masked by other processes, such as e.g. electronic/trigger noise that may dominate the event rate at very low energies. In this view, the very high displacement threshold energy of tungsten is interesting, as it results in the peak centered at $\sim 50$ eV, almost twice that of diamond.

\begin{figure}
	\centering
		\includegraphics{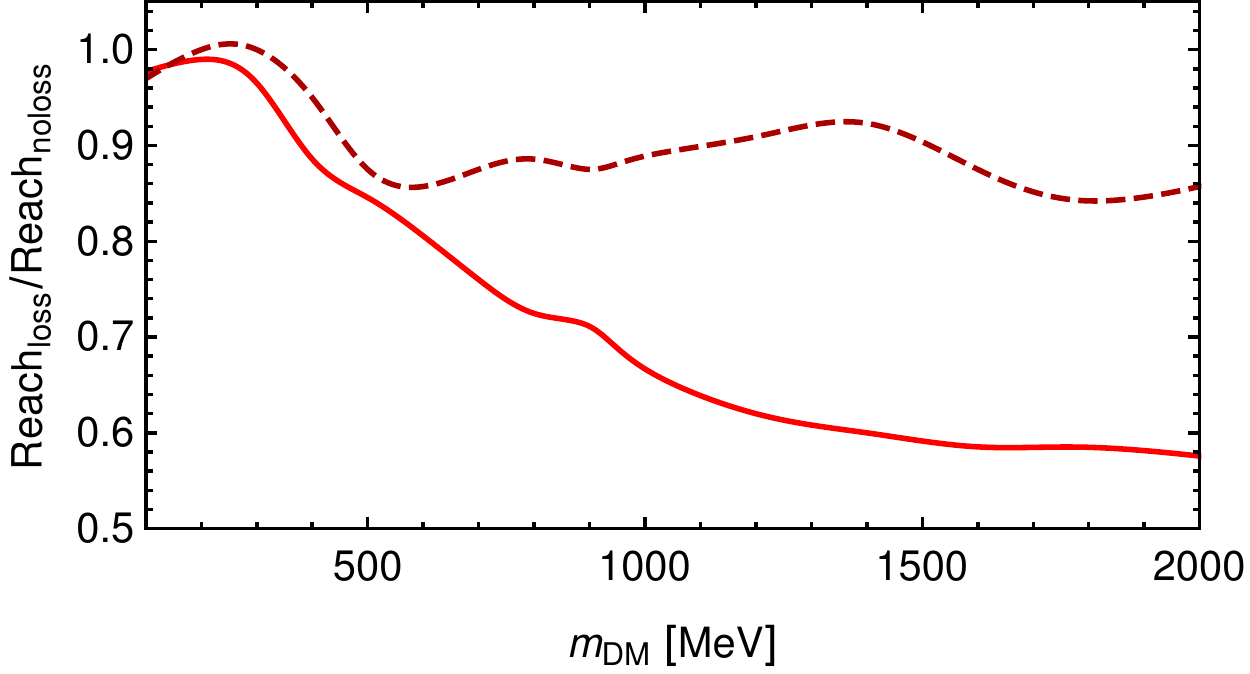}
		\caption{Gain in the reach of a diamond detector due to the energy loss effect against an electron recoil background with flat energy spectrum. The dashed line shows the ratio of the discovery reach (the lowest observable SI cross section) after the energy loss effect to the reach without the effect for a 1 kg year experiment. The solid line shows the same for a 100 kg year experiment.}
		\label{reach_ratio}
	\end{figure}

Because the peak feature is not expected for electron recoils, it can be used to enhance the signal identification for electron recoil dominated background. The detailed characterization of this effect will depend on the details of the experimental setup, but to get an idea of the possible gain, we have simulated an electron recoil background following the tritium background component reported in~\cite{SuperCDMS:2016wui}, and computed the discovery reach of a diamond detector for a spin-independent DM signal with or without the energy loss effect. To find the discovery reach we perform a likelihood-ratio test of the signal + background model to a background only model for the simulated data, where the signal model is the spin-independent DM spectrum with or without the energy loss. Figure \ref{reach_ratio} shows the gain, i.e. the ratio of the reach after the energy loss effect to the reach without the effect. For a 100 kg year experiment we observe a gain of $\mathcal{O}(40\%)$, and for a 1 kg year experiment the gain is $\mathcal{O}(10\%)$.

As a result of the motion of the lab frame with respect to the galactic rest frame, the dark matter recoil rate exhibits a preferred direction. The resulting daily modulation of the low energy event rate due to ionization threshold in ionization-based detectors has been studied in~\cite{Kadribasic:2017obi,Heikinheimo:2019lwg,Sassi:2021umf}. Due to the strong directional dependence of the energy loss in the crystal, one might wonder if a daily modulation effect should appear also in the peak feature shown in figure \ref{dRdE}. However, for the peak to appear in the spectrum, the DM particle must be heavy enough to excite recoils well above the defect creation threshold. As described in~\cite{Heikinheimo:2019lwg}, the angular recoil spectrum is then broad, so that the effect of the directional variation in the energy loss gets effectively averaged over, and there is no noticeable daily modulation in this effect.

\section{Conclusions}
\label{sec:checkout}
We have performed molecular dynamics simulations to determine the amount of energy lost in lattice deformations in low energy nuclear recoil events as a function of the recoil direction and energy. We have described how this effect modifies the observed recoil spectrum in phonon-based detectors. Some materials, most notably diamond, exhibit a sharp defect creation threshold, and for those materials the energy loss effect results in a prominent peak in the observed recoil spectrum. Via simulations of a number of materials with different physical properties, we have demonstrated a large variability in the profile of the energy loss and of the recoil spectrum.

In \cite{Heikinheimo:2021syx} we have discussed how the energy loss features can be used to determine whether the observed low energy excess event rate~\cite{Proceedings:2022hmu} is due to nuclear recoils. In this paper we have shown how the expected DM recoil rate changes as a result of the energy loss and depending on the target material, under the standard assumptions of a spin-independent interaction and standard halo model. It is evident from the results shown above that an ideal target material for observation of this effect is hard with simple crystal structure, and low propensity for damage recombination. Of the materials we have simulated, the most promising candidates are diamond at low recoil energies, and tungsten carbide at slightly higher recoil energies. The simulated results in this paper can be experimentally validated using low energy neutron beam wherein a monoenergetic neutron beam scatters off the detector and the recoil energy is measured using the scattering angle and/or time of flight like TUNL\footnote{\url{https://tunl.duke.edu}}.

The energy loss data described in this work can be found in an online repository\footnote{\url{https://github.com/sebsassi/elosssim}}, and we plan to add data for other detector materials in the future. In future work we shall also investigate how the peak feature is affected by relaxing the assumptions on the spin-independent interaction or on the DM halo model.

\section{Acknowledgements}

The authors wish to thank the Finnish Computing Competence Infrastructure (FCCI) (persistent identifier urn:nbn:fi:research-infras-2016072533) for supporting this project with computational and data storage resources. The financial support from Academy of Finland, project $\# 342777$, is gratefully acknowledged.

\bibliographystyle{hieeetr}	
\bibliography{main.bib}


\end{document}